\documentclass[nofootinbib,floats,aps,superscriptaddress,preprint]{revtex4} 
 
\newcommand{\beqa}{\begin{eqnarray}} 
\newcommand{\eeqa}{\end{eqnarray}} 
\newcommand{\beq}{\begin{equation}} 
\newcommand{\eeq}{\end{equation}} 
 
\newcommand{\pkl}{\mbox{$B\to \psi K_L$}} 
\newcommand{\pks}{\mbox{$B\to \psi K_S$}} 
\newcommand{\apks}{\mbox{$a_{\rm CP}(B\to \psi K_S)$}} 
\newcommand{\apkl}{\mbox{$a_{\rm CP}(B\to \psi K_L)$}} 
\renewcommand{\Im}{\mbox{Im}} 
\renewcommand{\Re}{\mbox{Re}}  
\newcommand{\reeps}{\mbox{$\Re\, \epsilon_K$}}  
\newcommand{\imeps}{\mbox{$\Im\, \epsilon_K$}}  
  
\newcommand{\lams}{\lambda_{\psi K^*}}  
\newcommand{\lamin}{\lambda^{-1}_{\psi \Kbar^*}}  
  
\def\Bbar{\,\overline{\!B}{}}  
\def\Kbar{\,\overline{\!K}{}}  
\begin{document}  
  
\preprint{\vbox{\hbox{LBNL--49680}\hbox{hep-ph/0204212}\hbox{April, 2002}}}  
  
\vspace*{1cm}  
  
\title{\boldmath Can the CP asymmetries in $\pks$ and $\pkl$ differ?}  
  
\author{Yuval Grossman}  
\affiliation{Department of Physics,  
Technion--Israel Institute of Technology,\\  
       Technion City, 32000 Haifa, Israel\vspace{6pt}}  
  
\author{Alexander L.\ Kagan}\thanks{On leave of absence from   
the Dept.\ of Physics, University of Cincinnati until Sep.\ 2002}   
\affiliation{Theory Group,  
	Fermi National Accelerator Laboratory,  
	Batavia IL 60510\vspace{6pt} }  
  
\author{Zoltan Ligeti\,}  
\affiliation{Ernest Orlando Lawrence Berkeley National Laboratory,  
        University of California, Berkeley CA 94720  
	\\[20pt] $\phantom{}$ }

\begin{abstract} \vspace*{8pt}  
  
In the standard model the CP asymmetries in $\pks$ and $\pkl$ are equal in 
magnitude and opposite in sign to very good approximation.  We compute the 
order $\epsilon_K$  corrections to each of these CP asymmetries and find that 
they give a deviation from $\sin 2\beta$ at the half percent level, which may 
eventually be measurable.  However, the correction to $\apks + \apkl$ due to 
$\epsilon_K$ is further suppressed. The dominant corrections to this sum, at 
the few times $10^{-3}$ level, come from the $B$ lifetime difference, and CP 
violation in $B-\Bbar$ mixing and $B \to \psi K $ decay. New physics could 
induce a significant difference in the $\sin(\Delta m_B\, t)$ time dependence 
in the asymmetries if and only if the ``wrong-flavor" amplitudes $B \to \psi 
\Kbar$ or $\Bbar \to \psi K$ are generated.  A scale of new physics that lies 
well below the weak scale would be required.  
Potential scenarios are therefore 
highly constrained, and do not appear feasible.  A direct test is proposed to 
set bounds on such effects. 
 
\end{abstract}  
  
\maketitle

\section{Introduction}

In the standard model (SM) the CP asymmetries in $\pks$ and $\pkl$ have the  
same magnitudes and opposite signs,  
\beq\label{known}  
\apks = -\apkl \,.  
\eeq  
Since these two modes have the largest weight in the BABAR and BELLE 
measurements of CP violation quoted as $\sin2\beta$~\cite{s2b}, it is important 
to understand the accuracy of Eq.~(\ref{known}) in the SM, and whether it could 
be altered by new physics.  In Section~II, we review the necessary formalism 
and explain the conditions that have to be fulfilled in order to violate 
Eq.~(\ref{known}), in the limit where the $K_S$ and $K_L$ are considered to be 
pure CP eigenstates. Specifically, we are interested in how different 
magnitudes for the $\sin(\Delta m_B\, t)$ terms in the asymmetries could be 
realized.  A necessary condition is shown to be the presence of ``wrong-flavor" 
kaon amplitudes, $B \to \psi \Kbar$ or $\Bbar \to \psi K$, which are negligibly 
small in the SM. 
 
There are corrections to both sides of Eq.~(\ref{known}) proportional to 
$\epsilon_K$, due to the fact that the $K_S$ and the $K_L$ are not pure CP 
eigenstates.  One may also expect the $K_S$ to $K_L$ lifetime ratio to enter, 
since the $K_S$ is identified experimentally by two pions that are produced at 
a distance from the interaction point that is less than a few times the typical 
$K_S$ decay length.  The probability that a $K_L$ decays into two pions within 
the same region is suppressed.  However, to obtain the corrections to 
Eq.~(\ref{known}), it is necessary to fully take into account interference 
effects between the (unobserved) intermediate $K$ and $\Kbar$ states.  In 
Section~III we show that $\apks$ and $\apkl$ receive corrections at order 
$\epsilon_K$, but the correction to Eq.~(\ref{known}) is further suppressed. 
  
In Section IV we investigate how new physics could yield the wrong-flavor kaon 
amplitudes required to obtain $\apks \ne - \apkl$. Sizable effects are possible 
in principle, but we find that the scale of new physics would have to lie well 
below the weak scale. Potential scenarios are therefore tightly constrained by 
bounds on flavor changing neutral current processes, and a significant 
contribution appears rather unlikely.  This is illustrated with an example that 
arises in supersymmetric models with an ultra-light sbottom squark.  We also 
discuss experimentally testable predictions which can be used to set bounds on 
the  wrong-flavor amplitudes. Section~V contains our conclusions.

\section{Formalism}

The time dependent CP asymmetries in $B\to \psi K_{S,L}$ (for notation 
and formalism, see~\cite{rev,Silvabook,fnalb}) are given by 
\beqa\label{aCP}  
a_{\rm CP}(B\to \psi K_{S,L})   
&=& {\Gamma(\Bbar(t) \to \psi K_{S,L}) -\Gamma( B(t) \to \psi K_{S,L}) \over  
  \Gamma(\Bbar(t) \to \psi K_{S,L}) +\Gamma( B(t) \to \psi K_{S,L})}   
  \nonumber\\[4pt]  
&=& -{(1 - |\lambda_{S,L}|^2) \cos(\Delta m_B\, t)   
  - 2 \Im\lambda_{S,L} \sin(\Delta m_B\, t) \over  1 + |\lambda_{S,L}|^2 }  
  \nonumber\\[2pt]  
&\equiv& S_{S,L}\, s_B - C_{S,L}\, c_B \,.  
\eeqa  
Here $s_B \equiv \sin(\Delta m_B\, t)$, $c_B \equiv \cos(\Delta m_B\,t)$,  
$\Delta m_B \equiv m_H - m_L $, 
and the last line defines $S_{S,L}$ and $C_{S,L}$.  Furthermore, 
\beq\label{lambdadef}  
\lambda_{S,L} \equiv \left({q_B \over p_B }\right)  
  \left({\bar A_{S,L} \over A_{S,L}}\right) ,  
\eeq  
where $\bar A_{S,L} \equiv A(\Bbar \to \psi K_{S,L})$ and  
$A_{S,L} \equiv A(B \to \psi K_{S,L})$.  
The neutral $B$ and $K$ meson mass eigenstates are defined in  
terms of flavor eigenstates as  
\beq  
|B_{L,H}\rangle = p_B |B \rangle \pm q_B |\Bbar \rangle \,, \qquad   
  |K_{S,L}\rangle = p_K|K \rangle \pm q_K |\Kbar \rangle \,.  
\eeq  
In the $|\lambda_{S,L} | = 1$ limit, which is usually considered, the  
asymmetries reduce to the simple form $a_{\rm CP}(B\to \psi K_{S,L}) = 
\Im\lambda_{S,L}\, \sin(\Delta m_B\, t)$, and $\Im\lambda_{S,L} = 
S_{S,L} = \pm \sin2\beta$ and $C_{S,L} = 0$.  Our goal is to 
investigate possible deviations from this limit. 
  
Since $B$ meson decays are better described in terms of flavor eigenstates at  
short distances, we rewrite $\lambda_{S,L}$ in terms of the right-flavor kaon  
decay amplitudes  
\beq  
\bar A_{\Kbar} \equiv A(\Bbar \to \psi \Kbar)\,, \qquad  
  A_K \equiv A(B \to \psi K)\,,   
\eeq  
and the wrong-flavor kaon decay amplitudes   
\beq  
\bar A_K \equiv A(\Bbar \to \psi K)\,, \qquad   
  A_{\Kbar} \equiv A(B \to \psi \Kbar )\,.  
\eeq  
To parameterize the contributions due to possible wrong-flavor amplitudes  
from new physics, we define  
\beq \label{abdef}  
a \equiv \left({q_K \over p_K}\right)   
  \left({\bar A_K \over \bar A_{\Kbar}}\right) , \qquad  
b \equiv \left({p_K \over q_K}\right) \Bigg({A_{\Kbar} \over  A_K}\Bigg) \,.  
\eeq  
Then we can rewrite $\lambda_{S,L}$ defined in Eq.~(\ref{lambdadef}) as  
\beq\label{main}  
\lambda_{S,L} = \pm \lambda_B \left( {1\pm a \over 1\pm b}\right) ,  
\eeq  
where \cite{Silvabook,Silva-rec} 
\beq\label{lambdabdef}  
\lambda_B \equiv \left({q_B \over p_B}\right)  
  \left({\bar A_{\Kbar} \over A_K}\right) \left({p_K \over q_K}\right) .  
\eeq  
In the SM, and in any extensions of it in which the wrong-flavor kaon 
amplitudes are negligibly small (i.e., $|a|$ and $|b| \ll 1$), Eq.~(\ref{main}) 
reduces to $\lambda_{S,L} = \pm \lambda_B$, and so $\lambda_S + \lambda_L = 
0$.  As a result, for the two CP asymmetries in Eq.~(\ref{aCP}), $S_S = - S_L$ 
and $C_S = C_L$ .  However, for arbitrary $a$ and $b$, 
\beq \label{const}  
\lambda_S+\lambda_L = \lambda_B\, {2(a-b) \over 1- b^2}\,.  
\eeq  
We learn that a necessary and sufficient condition for $\lambda_S \ne 
-\lambda_L$ is the presence of non-vanishing wrong-flavor amplitudes with $a 
\ne b$.  Such a situation can arise either if $|a| \ne |b|$ or if $\arg(a) \neq 
\arg(b)$. 
 
To get a rough idea of the size of the expected difference between the two  
asymmetries, note that if each right-flavor and wrong-flavor kaon amplitude is  
dominated by a single contribution, then $|a|\approx |b|$ holds.  We further  
assume that the CP violating phases are not small, namely that  
$\Re\lambda_{S,L} \sim \Im \lambda_{S,L} \sim {\cal O}(1)$ as in the SM, and  
that $\arg(a) \sim \arg(b) \sim {\cal O}(1)$.  Under these assumptions,  
\beq  
  \Im (\lambda_S + \lambda_L ) \sim |a|\,.  
\eeq    
Thus, $S_S + S_L$ is expected to be of the order of the ratio of wrong-flavor 
to right-flavor kaon amplitudes.  If the strong phases between the wrong-flavor 
and right-flavor kaon amplitudes are not small, effects of similar order will 
also be generated for the $C_{S,L}$ terms in the CP asymmetries in 
Eq.~(\ref{aCP}).

\section{The difference in CP asymmetries in the SM}

The amplitudes for $B$ decays to wrong-flavor kaons are negligible in the SM.   
Any contribution would require at least two $W$ propagators in an (exchange)  
annihilation graph,  and would involve small CKM matrix elements.  Naive  
estimates in the SM lead to $|a|,\ |b| < 10^{-6}$.   There are  much larger 
effects which contribute to the CP asymmetries at the $10^{-3}$  level.  Since 
they are all small, we can expand to linear order in  each of them. The finite 
$B$ meson width difference results in  equal contributions to the $S_S$ and 
$S_L$ terms in Eq.~(\ref{aCP}) ~\cite{DFN,fnalb,DHKY}.  The deviation of 
$|\lambda_B |$ from unity due to CP  violation in $B$ mixing or in $B \to \psi 
K$ decay results in non-zero $C_{S,L}$ terms,  satisfying $C_S = C_L$.  CP 
violation in decay also results in corrections to the $S_{S,L}$ terms of equal 
magnitude, but of opposite sign. We will return to a discussion of these 
effects later. 
 
CP violation in $K-\Kbar$ mixing contributes to $a_{\rm{CP}} (B \to \psi 
K_{S,L} )$ via corrections to $\lambda_B $.  If the measured final states were 
the $K_S$ and $K_L$ mass eigenstates, this would be the only effect of 
$\epsilon_K$ \cite{xing} and the relation $\lambda_S =-\lambda_L$ would not be
altered.   However, there is an additional effect due to the fact that the
experimentally  reconstructed $\psi K_S$ final state is actually a coherent
superposition of  $\psi K \to \psi \pi \pi $ and $\psi \Kbar \to \psi\pi\pi$
with some constraint  on the kaon decay time.  For example, at BABAR and BELLE
the $K_S$ is  identified by requiring two pions in the tracking system.  This
requirement  selects kaons that decay after a short time.  Thus they are mainly
$K_S$, but  there is a small $K_L$ admixture, since the $K_L$ can also decay to
two pions.   Final states reconstructed as $K_L$, on the other hand, are
identified by hits  in the hadronic calorimeter.  This requires that the kaon
decay time must be  much longer than the $K_S$ lifetime, therefore such states
are pure $K_L$ to  very good accuracy. 

To study the effect of kaon mixing, it is most convenient to use the  cascade 
mixing formalism~\cite{Kayser0,Kayser,azimov,Silva,Silvabook}.  In  particular, 
to obtain the total $B \to \psi K \to \psi \pi \pi $  amplitude a coherent sum 
is performed over the physical $B$ and $K$  mass eigenstate contributions,  
\beqa\label{cascade}  
A (B \to \psi K \to \psi \pi \pi) &=& \sum_{M,N}   
 A(K_N \to \pi \pi ) e^{-i (m_{K_N} - i\Gamma_{K_N}/2) t_K } \nonumber\\  
&&\quad \times A (B_M \to \psi K_N ) e^{-i (m_{B_M} - i\Gamma_{B_M}/2) t }\,  
  \langle B_M | B \rangle \,,  
\eeqa  
where $M = H,L$ and $N = S,L$ are summed over, and $t_K$ is the time  between 
the formation and decay of the $K$ meson.    
 
We are interested in obtaining the corrections to the CP asymmetries due to  
$\epsilon_K$, so in the following we set $\Delta\Gamma_B=0$ but allow  for 
deviations of $|\lambda_B |$ from unity. The resulting decay rates  can be 
expressed as  
\beq\label{master}  
\Gamma[ B(\Bbar) \to \psi K \to \psi \pi \pi]  
\propto \left[e^{-\Gamma_S t_K} c_{11} + e^{-\Gamma_L t_K} c_{22}    
  + 2 e^{-(\Gamma_S+\Gamma_L) t_K/2} c_{12} \right] e^{-\Gamma_B t} \,.  
\eeq  
For the $c_{ij}$ coefficients, following Ref.~\cite{Silva}, we obtain  
\beqa  
c_{11} &=& |1+\lambda_K|^2\, \left\{ 1+|\lambda_B |^2 \mp 2 s_B\,   
           \Im \lambda_B  
\pm c_B (1 -|\lambda_B |^2 ) \right\} \,,\nonumber\\[2pt]  
c_{22} &=& |1-\lambda_K|^2\,\left\{ 1 +|\lambda_B |^2 \pm 2 s_B\,  
           \Im \lambda_B  
\pm c_B (1 -|\lambda_B |^2) \right\} \,, \nonumber \\[2pt]  
c_{12} &=& \pm \left\{ 2 \left(1-|\lambda_K|^2 \right)   
  (c_B c_K - s_B s_K\,\Re\lambda_B)  
  - 4 \,\Im\lambda_K\, (c_B s_K + s_B c_K\, \Re\lambda_B) \right\}  
  \nonumber \\*[2pt]  
&& -\left(1- |\lambda_B |^2 \right) \left\{\left(1 - |\lambda_K|^2 \right)  
(\pm c_B c_K - c_K )  + 2 \Im\lambda_K (s_K \mp c_B s_K ) \right\}\, ,  
\eeqa  
where the upper (lower) signs stand for decays of a $B$ ($\Bbar$) meson,  
\beq  
\lambda_K \equiv {q_K \over p_K } \,   
      {A(\Kbar \to \pi \pi ) \over A(K \to \pi \pi)}\,,  
\eeq  
and  
\beq  
s_K \equiv \sin\Delta m_K t_K\,, \qquad  
c_K \equiv \cos\Delta m_K t_K\,.   
\eeq  
The $c_{11}$ ($c_{22}$) term corresponds to decays of $K_S$ ($K_L$), and  
the $c_{12}$ term is due to the interference between them.  
  
To obtain corrections to the CP asymmetries due to $\epsilon_K$ we need the  
following relations, valid to leading order in $\epsilon_K $ ($\epsilon_K'$ is  
neglected throughout),  
\beq\label{lambdaKeps}  
\lambda_K = 1 - 2\epsilon_K \,,  
\eeq  
where $|\epsilon_K| \approx 2.28 \times 10^{-3}$ and 
\beq\label{relnsI}  
\imeps = x_K\, \reeps \bigg[ 1 + {\cal O}\bigg(\frac{\Gamma_L}{\Gamma_S}  
  \bigg) \bigg]\,,\qquad  
x_K \equiv \frac{2  \Delta m_K }{\Gamma_S + \Gamma_L } \approx 0.95 \,.  
\eeq  
As can be seen from Eqs.~(\ref{lambdabdef}) and (\ref{lambdaKeps}),   
$\lambda_B$ to leading order in $\epsilon_K$ is given by  
\beq\label{lambdaBshift}  
\lambda_B = - e^{-2i\beta}\, (1+2 \epsilon_K) \,,  
\eeq  
where $\beta$ is the usual angle of the unitarity triangle.    
  
What is experimentally called $\apks$ is obtained by integrating the rates in  
Eq.~(\ref{master}) with respect to $t_K$ from (almost) zero to some cutoff  
$t_{\rm{cut}}$ that depends on the experimental setup, and then forming the  
asymmetry defined in Eq.~(\ref{aCP}). Since this cutoff is much larger  than 
the $K_S$ lifetime (by about a factor of 10 at BABAR and  BELLE), it is a good 
approximation to perform the integrals over the terms proportional to $c_{11}$ 
and $c_{12}$ from zero to  infinity.  Using the above relations, we find to 
leading order in  
$\epsilon_K$,  
\beq\label{ksasym}  
\apks =  \Big[ \sin2\beta - 2 \imeps \cos2\beta \Big] s_B   
  - 2 \reeps\, c_B \,,  
\eeq  
where it is to be understood that $K_S$ stands for the state identified in the 
experiments as $K_S$.  The corrections to $\apkl$ are obtained from 
Eq.~(\ref{aCP}), taking into  account the correction to $\lambda_B$ of order 
$\epsilon_K$ given in  Eq.~(\ref{lambdaBshift}).  The result is  
\beq\label{klasym}  
\apkl =  - \Big[ \sin2\beta - 2\imeps \cos2\beta \Big] s_B   
  + 2 \reeps\, c_B \,.  
\eeq  
Remarkably, to leading order in $\epsilon_K$, the relation $\apks = -\apkl$ is 
maintained.   The terms with $s_B$ time dependence in Eqs.~(\ref{ksasym}) and 
(\ref{klasym}) originate from $\Im \lambda_B$ and its small deviation from 
$\sin2\beta$.  The third term in Eq.~(\ref{ksasym}) receives contributions from 
both the interference term $c_{12}$ (given by $-4 \reeps\, c_B$), and the 
correction due to $|\lambda_B| \neq 1$ in $c_{11}$ (given by $+2 \reeps\, 
c_B$).  The relation  $\apks = -\apkl$  is maintained because the ratio of 
these two terms is $-2$, and the third term in Eq.~(\ref{klasym}) comes 
entirely from the $|\lambda_B| \neq 1$ contribution (that is $+2 \reeps\, 
c_B$). 
 
Corrections to $\apks = -\apkl$ due to $\epsilon_K \neq 0$ only occur 
suppressed by other factors, and are therefore not shown explicitly in 
Eqs.~(\ref{ksasym}) and (\ref{klasym}). There are contributions of order 
$\epsilon_K$ to $S_S$ from the $c_{12}$ interference term, which are suppressed 
by either $\imeps - x_K\, \reeps \propto \Gamma_L/\Gamma_S$ according to 
Eq.~(\ref{relnsI}), or by $e^{-\Gamma_S t_{\rm cut}/2}$ due to the finite 
experimental cut $t_K < t_{\rm cut}$.  The largest correction numerically 
actually comes from a contribution of the $c_{22}$ term to $S_S$, which is  
given by $-2|\epsilon_K|^2\, (1 -e^{- \Gamma_L t_{\rm cut}})\, \Gamma_S 
/\Gamma_L $.  For $t_{\rm cut} \sim 10\, \tau_S$, it is about $-1 \times 
10^{-4}$.   
 
To close this section, we return to discuss the relative importance of the 
corrections to the CP asymmetries from $\epsilon_K$, from the $B$ lifetime 
difference, and from CP violation in $B$ mixing and decay. The $B$ lifetime 
difference, $\Delta\Gamma_B  \equiv \Gamma_H - \Gamma_L $, modifies the 
asymmetries to first order in $\Delta\Gamma_B/\Gamma_B$ 
as~\cite{fnalb} 
\beq\label{delwidth}  
\delta a_{\rm CP}(B\to \psi K_{S,L})   
  = \frac12\, \sin 2\beta\, \cos2\beta\, (\Delta\Gamma_B\, t)\, s_B \,.  
\eeq  
Using $t \sim 1/\Gamma_B$ and the estimate $\Delta\Gamma_B / \Gamma_B \sim 3 
\times 10^{-3}$~\cite{DHKY}, these corrections are expected to be comparable to 
the $s_B$ terms arising at ${\cal O}(\epsilon_K)$.  (Note that new physics 
contributions to the $B$ lifetime difference  are unlikely to be sufficiently 
large to significantly modify the size of this effect.) Corrections due to CP 
violation in $B-\Bbar$ mixing ($|q_B/p_B| \neq 1$), to first order in 
$\Gamma_{12}/M_{12}$, only modify the $C_{S,L}$ terms in the asymmetries, and 
are given by 
\beq\label{delmix}  
\delta a_{\rm CP}(B\to \psi K_{S,L}) = -{1-|\lambda_B|^2 \over 2}\, c_B \,.  
\eeq  
At this order, $1 - |\lambda_B |^2 = \Im (\Gamma_{12}/M_{12})$, which is also 
equal to the measurable CP asymmetry in semileptonic decays, $A_{\rm SL}$. A 
recent estimate gives $\Im (\Gamma_{12}/M_{12} ) \approx - (0.5 - 1.3) \times 
10^{-3 }$~\cite{LLNP}, so this correction is somewhat smaller than the $c_B$ 
terms induced at  ${\cal O}(\epsilon_K )$.  Finally, we consider the effect of 
direct CP violation in $B\to \psi K_{S,L}$ decays due to the CKM suppressed 
penguin diagrams.  We denote by $T$ all contributions to the decay amplitude 
proportional to CKM elements $\lambda_c$  and by $P$ all contributions 
proportional to $\lambda_u$, where $\lambda_q \equiv V_{qb} V_{qs}^*$.  Then 
$A(\Bbar^0 \to \psi \Kbar^0) = \lambda_c\, T + \lambda_u P$, and the resulting 
modifications of the CP asymmetries are 
\beq\label{deldirect}  
\delta a_{\rm CP}(B\to \psi K_{S,L})  
= \mp 2 \cos2\beta\, \Im \frac{\lambda_u}{\lambda_c }\, \Re \frac{P}{T}\, s_B  
  - 2\, \Im \frac{\lambda_u}{\lambda_c}\, \Im \frac{P}{T}\, c_B \,.  
\eeq    
The CKM suppression ($|\lambda_u / \lambda_c |\sim  1/50$), and the hard to 
estimate matrix element suppression and strong phases in $P/T$ imply that such 
effects are of order a few times $10^{-3}$ or below. 
 
Note that Eqs.~(\ref{delwidth})--(\ref{deldirect}) include corrections which 
are of equal magnitude and sign for the two asymmetries.  Therefore, when 
combined with Eqs.~(\ref{ksasym}) and (\ref{klasym}), they introduce a 
difference between the magnitudes of $\apks$ and $\apkl$.  In view of the fact 
that $\apks$ may be measured below the percent level during the next decade, we 
collect Eqs.~(\ref{delwidth})--(\ref{deldirect}) and (\ref{ksasym}) to obtain 
\beqa\label{ksfinal}  
\apks &=& \left[ \sin2\beta - 2 \cos2\beta\, \imeps
  + \frac14 \sin4\beta\, (\Delta\Gamma_B\, t)  
  - 2 \cos2\beta\, \Im \frac{\lambda_u}{\lambda_c }\, \Re \frac{P}{T}  
  \right] s_B  \nonumber\\*  
&-& \left[ 2\, \reeps + \frac12\, \Im \bigg({\Gamma_{12}\over M_{12}}\bigg)   
  + 2\, \Im \frac{\lambda_u}{\lambda_c}\, \Im \frac{P}{T} \right] c_B \,.  
\eeqa 
  
We conclude that in the SM the $S_S = -S_L = \sin2\beta$ and $C_S = C_L = 0$ 
relations between the CP asymmetries in $\pks$ and $\pkl$ hold at the $1\%$ 
level, therefore it is safe to average the asymmetries in the $\pks$ and $\pkl$ 
modes.  However, below the percent level, there are several effects shown in 
Eq.~(\ref{ksfinal}) which can be calculated with varying degrees of reliability 
that enter the relation between $\apks$ and $\sin2\beta$.

  
\section{Constraints on New Physics}

Consider $A_{\Kbar}$, the amplitude of the wrong-flavor decay $B^0 \to \psi  
\Kbar$. As the final state does not contain a $\bar d$ quark, the decay must  
proceed via annihilation of the $B$ meson.  The flavor structure of the 
operator that mediates this decay is $( \bar d b)(\bar d s)( \bar c c)$.  
(Here, and in what follows, the color indices and Dirac structure of the 
operators are suppressed.)  While the $( \bar d b)( \bar d s)$ part, which 
violates flavor, must come from new short distance physics, the $( \bar c c)$ 
pair can be generated either by gluons or by exchange of heavy particles.  In 
the following we study both cases. 
 
First, we consider models where the $c \bar c$ pair is generated from 
gluon exchange.  The high energy theory is assumed to produce an 
effective four-Fermi interaction 
\beq  
O_4^{\rm NP} = {1 \over M_4^2}\, \bar d b\, \bar d s \,,  
\eeq  
where $M_4$ is the effective scale of new physics, which includes all possible  
dimensionless couplings.  The bounds on such operators are very strong, as we  
find below, so we may crudely estimate their contributions to $A_{\Kbar}$.    
The final state could be produced either by forming the $\psi$ in a color octet 
Fock state from a hard gluon, or via an OZI suppressed graph where the $\psi$ 
is formed out of three gluons.  Taking into account the fact that both 
processes  are power suppressed, and using factorization, we obtain 
\beq \label{O-4} 
A_{\Kbar} \lesssim {1 \over M_4^2\, m_B^2}\, {\Lambda_{\rm QCD} \over m_B}\, 
  \alpha_s f_B f_K f_\psi\, m_\psi\, (\epsilon_\psi \cdot p_K) \,.   
\eeq  
Upper bounds on such contributions to the amplitude can be obtained by 
considering the effect of the new operators on the rare decays $B^\pm \to K^\mp 
\pi^\pm \pi^\pm$~\cite{Singer} and $B^\pm \to \pi^\pm K_S $~\cite{trojan}. 
(In addition to modifying the $B^\pm \to \pi^\pm K_S $ decay rate, they must 
increase the ratio $\Gamma (B^\pm \to \pi^\pm K_S )/\Gamma (B^\pm \to \pi^0 
K^\pm )$ which cannot be much larger than its value in the SM according to 
current data.) Assuming factorization, the latter gives the strongest 
bound~\cite{trojan} 
\beq \label{bound-4}  
M_4 \gtrsim 3\, {\rm TeV} \,.  
\eeq  
For comparison, we note that in the SM the right-flavor amplitude using the  
factorization hypothesis is given by 
\beq\label{eq:SMAKfact}  
A_K^{\rm SM} = {G_F \over \sqrt 2}\, V_{cb}V_{cs}^*\, a_2\,  
  f_\psi\, m_\psi\, F_1\, (\epsilon_\psi \cdot p_K ) \,,  
\eeq  
where $F_1$ is the $B \to K$ form factor at $q^2 = m_\psi^2$, and $a_2$ depends 
on the current-current operators' Wilson coefficients, $C_1$ and $C_2$.  The 
observed $B \to \psi K_S$ rate is reproduced if $a_2\, F_1 \approx 0.2$ 
\cite{cheng}. Using $\alpha_s(m_b) = 0.2$ and $f_B \sim f_K \sim f_\psi \sim 
200\,$MeV, Eqs.~(\ref{O-4})--(\ref{eq:SMAKfact}) imply $|a| \lesssim 
O(10^{-4})$. While these estimates are very crude, it is clear that large 
effects cannot occur. 
 
Next, we turn to the case of $c \bar c$ pair production by exchange of heavy 
particles.  The wrong-flavor amplitudes would be due to six-quark operators, 
with an effective Hamiltonian of the form 
\beq\label{eq:hamiltonian}  
O_6^{\rm NP} = {1 \over M_6^5}\, \bar d b\, \bar d s\, \bar c c \,,  
\eeq  
where $M_6$ is the effective scale of new physics, which includes all possible 
dimensionless couplings.  A crude estimate of the wrong-flavor amplitude using 
factorization yields 
\beq\label{eq:AKfact}  
A_{\Kbar} \sim  {1 \over M_6^5}\, m_B f_B f_K f_\psi\, m_\psi\,  
  (\epsilon_\psi \cdot p_K ) \,.  
\eeq  
Comparing Eqs.~(\ref{eq:AKfact}) and (\ref{eq:SMAKfact}) we find    
\beq  
|a| \sim \left({20\,{\rm GeV} \over M_6}\right)^5 \,.  
\eeq  
Thus, a difference of CP asymmetries greater than a percent for the $\sin  
\Delta m_B\, t$ terms, i.e., $|\apks| - |\apkl| \gtrsim 10^{-2}$, would 
require  a new physics scale that lies well below the weak scale. 
 
We know of only one new physics scenario which could in principle 
accommodate large wrong-flavor amplitudes: supersymmetric models with a light 
bottom squark of mass $2-5.5\,$GeV and a light gluino of mass 
$12-16\,$GeV~\cite{berger}. Such models have been proposed to enhance the $b$ 
quark production cross section at the Tevatron. Among the new operators which 
can arise at tree-level, there are several of the form $\bar d b\, \tilde b^* 
\tilde b$.  Stringent upper bounds on their coefficients have been obtained 
from  rare $B$ decays~\cite{sbotfcnc}. Interactions of the desired form in 
Eq.~(\ref{eq:hamiltonian}) would be generated from these operators if the 
$R$-parity violating Yukawa couplings mediating $\tilde b \to \bar c \bar d$ 
and $\tilde b \to \bar c \bar s$ decays were also present. Moreover, large 
wrong-flavor amplitudes could be generated if these couplings were of unit 
strength. However, an upper bound of order $10^{-5}$ on the product of these 
two couplings from box-graph contributions to $K - \Kbar$ mixing implies that 
the wrong-flavor kaon amplitudes must be negligibly small, i.e., $|a| \sim |b| 
\lesssim 10^{-5}$.  This example illustrates the difficulties any scenario with 
large wrong-flavor amplitudes  would face due to the requirement of a low mass 
scale for new flavor-changing  interactions. The possibility of significantly 
different CP asymmetries in  $\pks$ and $\pkl$ decays is therefore extremely 
unlikely.  
 
The  most direct test for such new physics effects is  provided by searching 
for the wrong-flavor decay $B \to \psi \Kbar^*$, by studying the time 
dependence of flavor tagged $B$ decays.  It is very likely that the matrix 
elements are similar in $B$ decays to $\psi K^*$ and $\psi K$, so the ratios of 
the wrong-flavor to the right-flavor decay amplitudes should be similar in the 
two cases. Although $\psi \Kbar^*$ is a vector-vector final state, and thus it 
is a mixture of CP even and odd components, this is not expected to yield a 
significant difference in the ratio of wrong-flavor to right-flavor decay 
amplitudes.  In the presence of  wrong-flavor amplitudes the time dependent 
rate is 
\beq\label{wstest} 
\Gamma[B(\Bbar) \to \psi K^*] \propto 1 + |\lams|^2   
  \pm \left[(1-|\lams|^2)\, c_B - 2\, \Im\lams\, s_B \right].  
\eeq  
Here $\lams$ is of the order of the wrong-flavor to right-flavor amplitude  
ratio, and the upper (lower) signs stand for decays of $B$ ($\Bbar$).  The time 
dependence for the $\Bbar(B) \to\psi \Kbar^*$ decay is obtained by  replacing 
$\lams$ by $\lamin$ and $\pm$ by $\mp$ in Eq.~(\ref{wstest}).  Fitting to these 
time dependences, the $B$  factories should be able to bound the magnitudes of 
the wrong-flavor  amplitudes, which constrains $|\apks + \apkl|$ using 
Eq.~(\ref{const}).

  
\section{Conclusions}

The $\pks$ and $\pkl$ decays are the golden modes for studying CP violation, 
since the hadronic uncertainties are below the $1\%$ level. We studied  
possible effects within the standard model and in the presence of new physics 
that can make the absolute values of the CP asymmetries in these two channels 
different.   We computed the corrections due to $\epsilon_K$, taking into 
account the way the $K_S$ and $K_L $ mesons are identified at the $B$ 
factories, and found that although $\epsilon_K$ induces corrections to each CP 
asymmetry at the few times $10^{-3}$ level, it only introduces a difference in 
their magnitudes at the $10^{-4}$ level.  Nevertheless, in the SM the 
difference in the absolute values of the two CP asymmetries is of order 
$10^{-3}$ due to the $B$ lifetime difference and CP violation in $B - \Bbar$ 
mixing and in $B \to \psi K$ decay.  These effects modify the relation between 
$\apks$ and $\sin2\beta$ as summarized in Eq.~(\ref{ksfinal}). 
 
New physics in $B-\Bbar$ mixing, which would modify $\apks$ and $\apkl$ while 
leaving their magnitudes equal, has been extensively studied.  Direct CP 
violation in $B \to \psi K$ decays, which would lead to contributions equal in 
magnitude and opposite [same] in sign for the $\sin(\Delta m_B\, t )$ 
[$\cos(\Delta m_B\, t )$] terms in the two asymmetries, has also been discussed 
previously. We investigated how new physics could violate $\apks = -\apkl$ via 
unequal magnitudes for the $\sin(\Delta m_B\, t )$ terms, and found that the 
presence of the wrong-flavor kaon amplitudes $B \to \psi \Kbar$ or $\Bbar \to 
\psi K$ are necessary to obtain significant effects, i.e., in excess of $1\%$. 
(Small effects are, in principle, possible via new physics contributions to  
the $B$ lifetime difference.)  This would require a scale for new physics which 
lies well below the weak scale, therefore the existence of a viable scenario is 
unlikely due to bounds on flavor changing neutral currents.  Using the current 
data sets, it should be possible for the $B$ factories to put tight bounds on 
the related wrong-flavor $\Bbar \to \psi K^*$ and $B \to \psi \Kbar^*$ decay 
amplitudes. 
 
While it is important to constrain the decay amplitudes to wrong-flavor kaons 
experimentally, unless the results indicate large new physics contributions, it 
is safe to combine the $\apks$ and $\apkl$ measurements.  If and when $\apks$ 
will be measured at or below the one percent level, it will become important to 
include the various subleading effects discussed in this paper.

\acknowledgments  
It is a pleasure to thank Gerhard Buchalla, Uli Nierste, Soeren Prell, 
Marie-Helene Schune, Vivek Sharma, Jo${\tilde {\rm a}}$o Silva, and Zack 
Sullivan for helpful discussions. We thank the CERN and SLAC theory groups 
for hospitality while portions of this work were done. 
Y.G.~was supported in part by the Israel Science Foundation under Grant 
No.~237/01-1. A.K. was supported in part by the Department of Energy under 
Grant DE-FG02-84ER40153. Z.L.~was supported in part by the Director, Office of 
Science, Office of High Energy and Nuclear Physics, Division of High Energy 
Physics, of the U.S.\ Department of Energy under Contract DE-AC03-76SF00098.  
The work of Y.G.\ and Z.L.\ was also supported in part by the United 
States--Israel Binational Science Foundation (BSF) through Grant No. 2000133.


\end{document}